\documentclass[twocolumn,showpacs,preprintnumbers,amsmath,amssymb,superscriptaddress]{revtex4}
\usepackage{graphicx,dcolumn,bm,mathrsfs}

\newcommand{\Eq}[1]{Eq.\,(\ref{#1})}
\newcommand{\Eqs}[1]{Eqs.\,(\ref{#1})}
\newcommand{\Fig}[1]{Fig.\,\ref{#1}}
\newcommand{\Figs}[1]{Figs.\,\ref{#1}}
\newcommand{\be}{\begin{equation}}
\newcommand{\ee}{\end{equation}}
\newcommand{\bea}{\begin{eqnarray}}
\newcommand{\eea}{\end{eqnarray}}
\newcommand{\bsube}{\begin{subequations}}
\newcommand{\esube}{\end{subequations}}
\newcommand{\la}{\langle}
\newcommand{\ra}{\rangle}
\newcommand{\lla}{\langle\!\langle}
\newcommand{\rra}{\rangle\!\rangle}

\begin{document}

\title{Spin-resolved counting statistics as a sensitive probe of spin correlation in \\ transport through a quantum dot spin valve}

\author{Guanjian Hu}
\affiliation{Department of Physics, Zhejiang University of Science and Technology, Hangzhou 310023, China}

\author{Shikuan Wang}
\affiliation{Department of Physics, Hangzhou Dianzi University, Hangzhou 310018, China}

\author{Jing Hu}
\affiliation{Department of Physics, Zhejiang University of Science and Technology, Hangzhou 310023, China}

\author{RuiQiang Li}
\affiliation{Department of Physics, Zhejiang University of Science and Technology, Hangzhou 310023, China}

\author{Yiying Yan}
\affiliation{Department of Physics, Zhejiang University of Science and Technology, Hangzhou 310023, China}

\author{JunYan Luo}
\email{jyluo@zust.edu.cn}
\affiliation{Department of Physics, Zhejiang University of Science and Technology, Hangzhou 310023, China}

\begin{abstract}
We investigate the noise in spin transport through a single quantum dot (QD) tunnel coupled to
ferromagnetic electrodes with noncollinear magnetizations.
Based on a spin-resolved quantum master equation, auto- and cross-correlations of spin-resolved currents are
analyzed to reveal the underlying spin transport dynamics and characteristics for various polarizations.
We find the currents of majority and minority spins could be strongly autocorrelated despite uncorrelated charge transfer.
The interplay between tunnel coupling and the Coulomb interaction gives rise to an exchange magnetic
field, leading to the precession of the accumulated spin in the QD. It strongly suppresses the bunching of spin tunneling events and results in a unique double-peak structure in the noise of the net spin current.
The spin autocorrelation is found to be susceptible to magnetization alignments, which may serve as a sensitive tool to measure the magnetization directions between the ferromagnetic electrodes.
\end{abstract}
\maketitle

\section{\label{introduction}Introduction}

The manipulation of the spin degrees of freedom lies at the heart of spintronics and spin-based information
processing \cite{Zut04323,bad1071,Jan12400,Die14187,Lin15307,Hof15047001,Bal18015005,Die20446,Yak202557,Ats20166711}.
Significant efforts have been devoted to explore new spin-based functional devices with high performance and
efficiency \cite{Sah0599,Le1182,Spa19203,Yan203232}.
Among them, a nanoscale spin valve is widely considered as an essential candidate building block in spintronic
devices \cite{Wey05115334,Gor05116806,Cot06235316,Bra04195345,Wey07155308,Par06144420,Mun20045404,Sto221,Wan22094416,Wan231,Tse211}.
It is of great importance to investigate its transport properties, where the spin-resolved correlations have a vital role to play.

Spin valves could be constructed in layered heterostructures, which are typically characterized by the presence of a strong tunnel magnetoresistance effect \cite{Par04862,Ike08082508}.
In recent years, these systems have especially benefited from the progress of discovering suitable 2D magnetic materials \cite{Son181214,Gon17265,Hua17270}.
A quantum dot (QD) system is another ideal platform for spin valve devices, with the unique advantage of precise manipulation and control of individual spins.
This is attributed to the rapid development of nanofabrication, which enables accurate confinement of single electrons and their accumulated spins in QDs \cite{Pet052180,Kop06766,Now071430,Ben22429,Ren224216,Ren22076802,Zho21235417} due to Coulomb blockade and spin blockade, respectively.

In the pioneering experiments, spin blockade takes place due to a blocking triplet state \cite{Mur07035432} in a double quantum dot \cite{Han071217,Ono021313,Liu05161305,Joh05165308,Fra06205333}.
Instead, we investigate the intriguing nonequilibrium spin on a single QD spin valve, where the spin blockade mechanism is ascribed to the spin selection and filtering between spins in the QD and the noncollinearly polarized ferromagnetic (FM) electrodes.
The magnitude and direction of the accumulated spin on QD is determined by two processes.
First, the tunnel coupling to the FM electrodes gives rise to a competition mechanism between injection and decay of spin on the QD.
Second, the interplay between the Coulomb interaction inside the QD and the tunnel coupling to the FM electrodes gives rise to an exchange (effective) magnetic field \cite{Bal01035310,Wet05020407,Wet06224406,Kon03166602,Kol07348,Luo20125410}, leading to spin precessional dynamics inside the QD.
This results in a prominent negative differential conductance in the charge current \cite{Mat04195345,Luo1159,Luo11145301} and remarkable suppression of the low frequency charge current noise \cite{Mat06075328}.
However, the exchange field is reminiscent of spin torque in magnetic structures, which translates to a net spin
angular momentum transfer or a spin current between the QD and FM electrodes.
It is thus appealing to investigate the spin current noise characteristics, instead of its charge counterparts.

Spin current noise, due to the discreteness of the spin carrier,  is a measure of the correlations between spin transfer \cite{Ant04046601}. It is able to provide additional information about the spin dynamics and spin transfer processes, different from their charge behaviors.
Recent studies have shown that the spin current noise could be generated even in the absence of a net charge
current \cite{Don05066601,Luo17035154}.
Furthermore, spin current fluctuations are predicted to be able to sensitively explore the effect of spin-flip
scattering \cite{Mis03100409,San03214501,Mis04073305,Lam04081301,Bel04140407}, and intriguingly probe repulsive or attractive interactions in QD transport systems \cite{Sau04106601,Luo14892}.

In this work, we investigate exclusively the spin current and its noise correlations in transport through a spin-valve, where the setup is composed of a single QD sandwiched between two noncollinearly polarized FM electrodes, as schematically shown in \Fig{fig1}.
Based on a spin-resolved quantum master equation, we obtain individual spin-resolved currents, as well as their autocorrelations and cross-correlations.
Our analysis takes fully into account the interplay between spin injection and spin precessional dynamics.
In case of perpendicular alignment, it is found that even for low bias with almost an empty QD, the autocorrelations of the majority and minority spin currents exhibit opposite dependence on polarization, although the charge transfers independently.
At a medium bias, the exchange filed strongly suppresses the bunching of spin tunneling events and results in a unique double-peak structure in the net spin current noise.
Furthermore, the spin current autocorrelations show a striking difference for perpendicular and antiparallel alignments, which may serve as a sensitive tool to measure the magnetization directions between the two FM electrodes.

The rest of the paper is organized as follows. In Sec. \ref{model}, we introduce the single QD spin valve setup with the corresponding  Hamiltonian. It is then followed by the introduction of the spin-resolved full counting statistics (FCS) in Sec. \ref{full_counting_statistics}. Section \ref{results} is devoted to the discussion of spin-resolved currents and their noise for various polarizations and magnetization directions in the electrodes. Finally, we summarize our work in Sec. \ref{conclusion}.

\section{\label{model}Model System}

\begin{figure}
\includegraphics[scale=0.35]{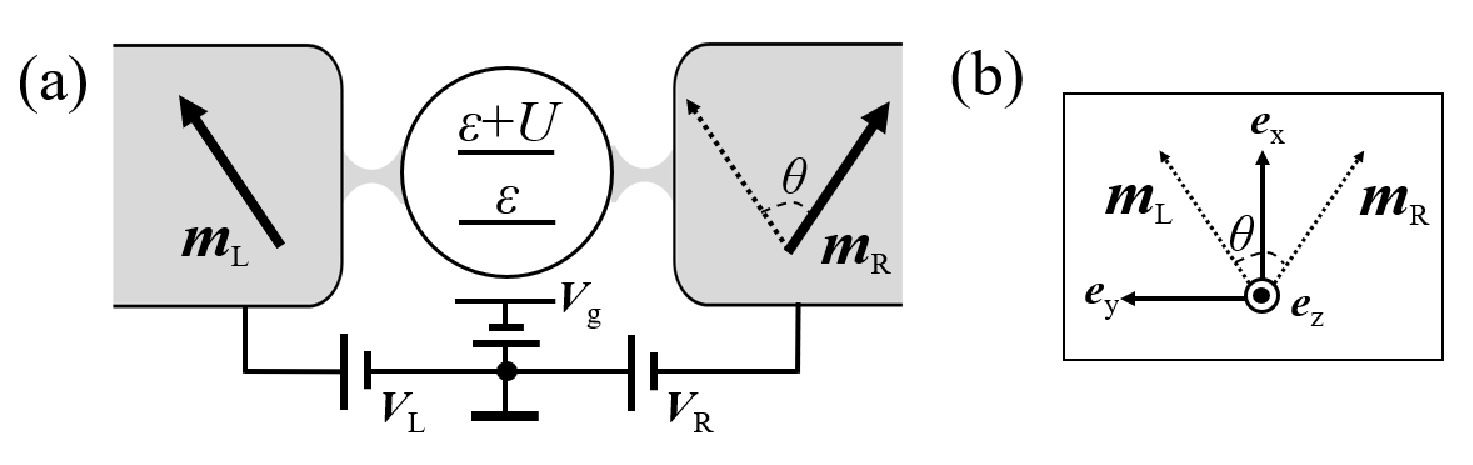}
\caption{\label{fig1} (a) Schematics of the QD spin valve system: A QD is tunnel coupled to the left and right FM electrodes,  whose  magnetizations directions are $\bm{m}_{\mathrm{L}}$ and $\bm{m}_{\mathrm{R}}$, respectively.
A bias voltage $V=V_{\mathrm{L}}-V_{\mathrm{R}}$ is applied accross the device, leading to spin and charge transport through the system.
(b)	The spin quantization axis of the QD, $\bm{e}_z$, which is chosen to be perpendicular to the plane spanned by $\bm{m}_{\mathrm{L}}$ and $\bm{m}_{\mathrm{R}}$ enclosing an angle $\theta\in[0,180^\circ]$.
The $\bm{e}_x$ and $\bm{e}_y$ are defined according to
$\bm{{\mathrm{e}}}_x\equiv\frac{\bm{m}_\mathrm{L}+\bm{m}_\mathrm{R}}{|\bm{m}_\mathrm{L}+\bm{m}_\mathrm{R}|}$ and $\bm{{\mathrm{e}}}_y\equiv\frac{\bm{m}_\mathrm{L}-\bm{m}_\mathrm{R}}{|\bm{m}_\mathrm{L}-\bm{m}_\mathrm{R}|}$, respectively.}
\end{figure}

The QD spin valve system is schematically shown in \Fig{fig1}, where the single QD is tunnel coupled to the left and right FM eletrodes. The Hamiltonian of the entire system $H_{\mathrm{Total}}$ reads
\begin{align}\label{Hamiltonian_total}
H_{\mathrm{Total}}=H_{\mathrm{L}}+H_{\mathrm{R}}+H_{\mathrm{QD}}+H_{\mathrm{TL}}+H_{\mathrm{TR}}.
\end{align}
The first two terms describe the left and right FM electrodes, which are modeled as reservoirs of noninteracting electrons.
According to the Stoner model of ferromagnetism\cite{Mar886949}, there is finite asymmetry in the density of
states for majority [${\cal D}_{\ell+}(\omega)$] and minority [${\cal D}_{\ell -}(\omega)$] spins in the electrode $\ell\in\{{\rm L,R}\}$.
Without loss of generality, the quantization axis for the electron spins in each electrode is chosen along the direction of the majority spins.
The magnetizations of the two FM electrodes are not necessarily to stay in the same direction and normally enclose an angle $\theta\sphericalangle (\bm{m}_{\rm L},\bm{m}_{\rm R})$, see \Figs{fig1}(a) and (b).
For simplicity, the densities of states are approximated to be energy independent in the following, i.e., ${\cal D}_{\ell\pm}(\omega)\to {\cal D}_{\ell\pm}$.
The degree of spin polarization thus can be characterized by $p_\ell=({\cal D}_{\ell+}-{\cal D}_{\ell -})/({\cal D}_{\ell +}+{\cal D}_{\ell -})$, where $p_\ell=0$ corresponds to
a nonmagnetic electrode, and $p_\ell=1$ a half-metallic electrode with majority spins only.
The corresponding Hamiltonian of the FM electrodes thus read
\begin{equation}
H_{\rm B}=\sum_{\ell={\rm L,R}}H_\ell=\sum_{\ell k\nu}\varepsilon_{\ell k\nu}a^{\dagger}_{\ell k\nu}a_{\ell k\nu},
\end{equation}
where $a_{\ell k\nu}$ ($a_{\ell k\nu}^{\dagger}$) is the annihilation (creation) operator for an electron with momentum $k$ of majority ($\nu=+$) or minority ($\nu=-$) spin in the FM electrode $\ell=\{{\rm L,R}\}$.
Each electrode is in thermal equilibrium and is characterized by the Fermi function $f_\ell(\omega)=\{1+e^{\beta_\ell (\omega-\mu_\ell)}\}^{-1}$, where
$\beta_\ell=(k_{\rm B}T_\ell)^{-1}$ is the inverse temperature and $\mu_\ell$ the chemical potential of the electrode $\ell$.
The difference in the chemical potentials defines the bias voltage across the two electrodes, i.e., $V=\mu_{\rm L}-\mu_{\rm R}$.
Hereafter, we choose $\hbar=e=1$, unless stated otherwise.

The Hamiltonian $H_{\rm QD}$ describes  the single QD, whose explicit form depends on the choice of the spin quantization axis. Here we chose neither $\bm{m}_{\rm L}$ nor $\bm{m}_{\rm R}$, but the $z$-axis perpendicular to the plane spanned by $\bm{m}_{\mathrm{L}}$ and $\bm{m}_{\mathrm{R}}$.
The unit vectors $\bm{e}_x$ and $\bm{e}_y$ are defined according to $\bm{e}_x\equiv\frac{\bm{m}_\mathrm{L}+\bm{m}_\mathrm{R}}{|\bm{m}_\mathrm{L}+\bm{m}_\mathrm{R}|}$ and $\bm{e}_y\equiv\frac{\bm{m}_\mathrm{L}-\bm{m}_\mathrm{R}}{|\bm{m}_\mathrm{L}-\bm{m}_\mathrm{R}|}$, respectively, as shown in \Fig{fig1}(b).
The Hamiltonian of the single QD thus reads
\begin{align}
H_{\mathrm{QD}}=\sum_{\sigma=\uparrow,\downarrow}\varepsilon c^{\dagger}_{\sigma}c_{\sigma}+Uc^{\dagger}_{\uparrow}c_{\uparrow}c^{\dagger}_{\downarrow}c_{\downarrow},
\end{align}
where $c_{\sigma}$ ($c_{\sigma}^{\dagger}$) is the annihilation (creation) operator for an up- ($\sigma=\uparrow$) or a down-spin ($\sigma=\downarrow$) electron, $\varepsilon$ is the spin-degenerate energy level of the single QD, and $U$ is the Coulomb energy cost for double occupation.

Electron tunneling between the FM electrodes and QD is described by the tunneling Hamiltonians $H_{\rm TL}$  and $H_{\rm TR}$.
In the spin quantization axis as shown in \Fig{fig1}(b), $H_{\rm TL}$ is given by
\begin{align}\label{Hamiltonian_tunnel}
H_{\mathrm{TL}}=\frac{1}{\sqrt{2}} & \sum_k \biggl\{ t_{\rm L k+} a^{\dagger}_{\mathrm{L}k+}(e^{+\mathrm{i}\frac{\theta}{4}}c_{\uparrow}+e^{-\mathrm{i}\frac{\theta}{4}}c_{\downarrow})
\nonumber \\
+&t_{\rm L k-} a^{\dagger}_{\mathrm{L}k-}(-e^{+\mathrm{i}\frac{\theta}{4}}c_{\uparrow}+e^{-\mathrm{i}\frac{\theta}{4}}c_{\downarrow})\biggr\}
+\mathrm{H.c.},
\end{align}
where $t_{\rm L k \nu} $ is the tunneling amplitude between the QD and the left electrode.
The tunneling between the right electrode and the QD $H_{\mathrm{TR}}$ can be obtained simply by the replacement $\mathrm{L}\rightarrow\mathrm{R}$ and $\theta\rightarrow-\theta$.
The corresponding tunnel coupling strengths are characterized by the intrinsic tunneling widths
$\Gamma_{\ell\pm} (\omega)=2\pi \sum_k |t_{\ell k \pm}|^2\delta(\omega-\varepsilon_{\ell k \nu})$, which will be approximated to be energy independent in the wide band limit, i.e., $\Gamma_{\ell\pm}(\omega)=\Gamma_{\ell\pm}$.

\section{Spin-resolved Full Counting Statistics}
\label{full_counting_statistics}

To keep track of spin and charge transport between the QD and the FM electrodes, we utilize the powerful FCS and introduce a group of counting fields $\bm{\chi}=(\chi_{{\rm L}+},\chi_{{\rm L}-},\chi_{{\rm R}+},\chi_{{\rm R}-})$ associated with transfer of spin ``$+$'' or ``$-$'' through the left (L) or right (R) junctions.
In the Fock state of the QD: $|0\ra$, $|\uparrow\ra$, $|\downarrow\ra$, and $|{\rm d}\ra$, standing for no extra electron, an extra spin-$\uparrow$ electron, an extra spin-$\downarrow$ electron, and double occupation respectively, the reduced density matrix can be expressed in a column vector  $\bm{\rho}(\bm{\chi})=(\rho_{00},\rho_{\uparrow\uparrow},\rho_{\downarrow\downarrow},\rho_{\rm dd},\rho_{\uparrow\downarrow},\rho_{\downarrow\uparrow})^{\rm T}$, where the diagonal element $\rho_{\alpha\alpha}\equiv\la \alpha|\rho|\alpha\ra $ denotes the probability of finding the QD in the state $|\alpha\ra$ ($\alpha$=0, $\uparrow$,  $\downarrow$, ${\rm d}$), and the off-diagonal elements $\rho_{\uparrow\downarrow}=\rho^\ast_{\downarrow\uparrow}\equiv\la \uparrow|\rho|\downarrow\ra$ stand for the quantum coherences. The other off-diagonal elements between states with different electron numbers are dynamically decoupled and thus not included.

The central quantity of the FCS is the spin-resolved cumulant generating function (CGF) ${\cal F}(\bm{\chi},t)$, which is defined via \cite{Esp091665}
\begin{align}\label{CGF}
e^{{\cal F}(\bm{\chi},t)}=&{\rm tr}_{\rm S}[\rho(\bm{\chi},t)]
=\rho_{00}(\bm{\chi},t)+ \rho_{11}(\bm{\chi},t)+\rho_{\rm dd}(\bm{\chi},t),
\end{align}
where ${\rm tr}_{\rm S}[\cdots]$ stands for the trace over the QD degrees of freedom and $\rho_{11}(\bm{\chi},t)=\rho_{\uparrow\uparrow}(\bm{\chi},t)+\rho_{\downarrow\downarrow}(\bm{\chi},t)$ is the probability to find the QD occupied by one electron regradless of spin orientations.
With the knowledge of the spin-resolved CGF ${\cal F}(\bm{\chi},t)$, various spin-resolved cumulants can be evaluated simply by taking partial derivative with respect to the corresponding counting fields.
Equation (\ref{CGF}) motivates us to investigate the spin-resolved probabilities $\rho_{00}(\bm{\chi},t)$, $\rho_{11}(\bm{\chi},t)$, and $\rho_{\rm dd}(\bm{\chi},t)$. Under the usual second-order Born-Markov approximation, they are found to satisfy
\begin{subequations}
\label{master_equations_chi_average_spin}
\begin{widetext}
\begin{align}
\dot{\rho}_{00}
=&-2\sum_{\ell={\rm L,R}}\gamma^+_{\ell}\rho_{00}+\sum_{\ell={\rm L,R}}(K^+_{\ell+}+K^+_{\ell-})\gamma_{\ell}^-
\rho_{11}
\nonumber \\
&+2\sum_{\ell={\rm L,R}}(K^+_{\ell+}-K^+_{\ell-})\gamma_\ell^-\cos(\textstyle \frac{\theta}{2})S_x+2[(K^+_{\rm L+}-K^+_{\rm L-})\gamma_{\rm L}^--(K^+_{\rm R+}-K^+_{\rm R-})\gamma_{\rm R}^-]\sin(\textstyle \frac{\theta}{2})S_y,
\\
\dot{\rho}_{11}=&2\sum_{\ell={\rm L,R}}(K^-_{\ell+}+K^-_{\ell-})\gamma^+_{\ell}\rho_{00}-\sum_{\ell=\rm L,R}(\gamma_{\ell}^-+\tilde{\gamma}_{\ell}^+)\rho_{11}+2\sum_{\ell={\rm L,R}}({K}^+_{\ell+}+{K}^+_{\ell-})\tilde{\gamma}^-_{\ell}\rho_{\rm dd}
\nonumber \\
&+2\sum_{\ell=\rm L,R}p_\ell(\tilde{\gamma}_{\ell}^+-\gamma_{\ell}^-)\cos({\textstyle \frac{\theta}{2}})S_x+2[p_{\rm L}(\tilde{\gamma}_{\rm L}^+-\gamma_{\rm L}^-)-p_{\rm R}(\tilde{\gamma}_{\rm R}^+-\gamma_{\rm R}^-)]\sin({\textstyle \frac{\theta}{2}})S_y,
\\
\dot{\rho}_{\rm dd}=&\sum_{\ell={\rm L,R}}({K}^-_{\ell+}+{K}^-_{\ell-})\tilde{\gamma}^+_{\ell}\rho_{11}-2\sum_{\ell=\rm L,R}\tilde{\gamma}_{\ell}^-\rho_{\rm dd}-2\sum_{\ell={\rm L,R}}({K}^-_{\ell+}-{K}^-_{\ell-})\tilde{\gamma}^+_{\ell}\cos({\textstyle \frac{\theta}{2}})S_x
\nonumber \\
&-2[({K}^-_{\rm L+}-{K}^-_{\rm L-})\tilde{\gamma}_{\rm L}^+-({K}^-_{\rm R+}-{K}^-_{\rm R-})\tilde{\gamma}_{\rm R}^+]\sin({\textstyle \frac{\theta}{2}})S_y,
\end{align}
\end{widetext}
where we have introduced $K^{\pm}_{\ell\nu}=\kappa_{\ell\nu}e^{\pm i\chi_{\rm\ell\nu}}$ and  $\kappa_{\ell\pm}={\cal D}_{\ell\pm}/({\cal D}_{\ell+}+{\cal D}_{\ell-})$ for simplicity. Apparently, the probabilities are coupled to each other due to tunneling between the QD and the FM electrode, where the tunneling rates are given by  $\gamma_\ell^{\pm}=\Gamma_\ell f_\ell^{\pm}(\varepsilon_\ell)$ and $\tilde{\gamma}_\ell^{\pm}=\Gamma_\ell f_\ell^{\pm}(\varepsilon_\ell+U)$, with $f_\ell^+(\omega)=f_\ell(\omega)$ the usual Fermi function and $f_\ell^-(\omega)=1-f_\ell(\omega)$.
Furthermore, it is found that these probabilities are also coupled to the average spin on the QD
\[
S_x=\frac{\rho_{\uparrow\downarrow}+\rho_{\downarrow\uparrow}}{2},\ S_y=i\frac{\rho_{\uparrow\downarrow}-\rho_{\downarrow\uparrow}}{2},\
S_z=\frac{\rho_{\uparrow\uparrow}-\rho_{\downarrow\downarrow}}{2},
\]
which are described by
\begin{widetext}
\begin{align}
\dot{S}_x=&\sum_{\ell={\rm L,R}}(K^-_{\ell+}-K^-_{\ell-})\gamma^+_{\ell}\cos({\textstyle \frac{\theta}{2}})\rho_{00}+\frac{1}{2}\sum_{\ell=\rm L,R}p_\ell(\tilde{\gamma}_{\ell}^+-\gamma_{\ell}^-)\cos({\textstyle \frac{\theta}{2}})\rho_{11}
\nonumber \\
&-\sum_{\ell={\rm L,R}}(\tilde{K}^+_\ell-\tilde{K}^+_\ell)\tilde{\gamma}_\ell^-\cos({\textstyle \frac{\theta}{2}})\rho_{\rm dd}-\sum_{\ell=\rm L,R}(\tilde{\gamma}_{\ell}^++\gamma_{\ell}^{-})S_x-(B_{\rm L}-B_{\rm R})\sin({\textstyle \frac{\theta}{2}})S_z,\label{dotSa}
\\
\dot{S}_y=&[(K^+_{\rm L+}-K^+_{\rm L-})\gamma_{\rm L}^+-(K^+_{\rm R+}-K^+_{\rm R-})\gamma_{\rm R}^+]\sin({\textstyle \frac{\theta}{2}})\rho_{00}+\frac{1}{2}[p_{\rm L}(\tilde{\gamma}_{\rm L}^+-\gamma_{\rm L}^-)-p_{\rm R}(\tilde{\gamma}_{\rm R}^+-\gamma_{\rm R}^-)]\sin({\textstyle \frac{\theta}{2}})\rho_{11}
\nonumber \\
&-[({K}^+_{\rm L+}-{K}^+_{\rm L-})\tilde{\gamma}_{\rm L}^-+({K}^+_{\rm R+}-{K}^+_{\rm R-})\tilde{\gamma}_{\rm R}^-]\sin({\textstyle \frac{\theta}{2}})\rho_{\rm dd}-\sum_{\ell=\rm L,R}(\tilde{\gamma}_{\ell}^++\gamma_{\ell}^{-})S_y+(B_{\rm L}+B_{\rm R})\cos({\textstyle \frac{\theta}{2}})S_z,\label{dotSb}
\\
\dot{S}_z=&(B_{\rm L}-B_{\rm R})\sin({\textstyle \frac{\theta}{2}})S_x-(B_{\rm L}+B_{\rm R})\cos({\textstyle \frac{\theta}{2}})S_y-\sum_{\ell=\rm L,R}(\gamma_{\ell}^-+\tilde{\gamma}_{\ell}^+)S_z.\label{dotSc}
\end{align}
\end{widetext}
\end{subequations}
It is apparent that the average spin is coupled to the probabilities. These terms are responsible for the building up of spin in the QD.
However, the tunneling coupling also leads to a decay of the average spin, with the decay rate given by $\gamma_{\rm dec}=\sum_{\ell=\rm L,R}(\gamma_{\ell}^-+\tilde{\gamma}_{\ell}^+)$.
Intriguingly, the interplay between Coulomb interaction and tunel-coupling between the QD and the FM electrode $\ell=$\{L, R\} gives rise to an exchange magnetic field \cite{Jeo19024507,Kon03166602}
\begin{align}\label{Bell}
{\bm B}_\ell=\frac{p_\ell\Gamma_\ell}{\pi}\int' d\omega\biggl(\frac{f_\ell^+(\omega)}{\omega-\varepsilon_\ell-U}+\frac{f_\ell^-(\omega)}{\omega-\varepsilon_\ell}\biggr) {\bm m}_\ell,
\end{align}
where the prime at the integral stands for the Cauchy's principle value.
The total exchange magnetic field ${\bm B}=\bm{B}_{\rm L}+\bm{B}_{\rm R}$ leads to precession of the average spin in the QD.
According to \Eqs{dotSa}-(\ref{dotSc}), it is described by
\be
\dot{\bm S}=\bm{S}\times\bm{B}.
\ee
We will reveal the essential roles it plays in the spin-resolved transport properties.

It should be noted that in the steady state limit ($t\to\infty$), the spin-resolved CGF is reduced to \cite{Ant04046601,Luo20125410,Luo23125113}
\be
\mathcal{F}(\bm{\chi},t)=\lambda_0(\bm{\chi}) t,
\ee
where $\lambda_0(\bm{\chi})$ is the unique eigenvalue of the Liouvillian associated with the quantum master equation (\ref{master_equations_chi_average_spin}) that satisfies  $\lambda_0(\bm{\chi})|_{\bm{\chi}\to\bm{0}}\to0$.
For instance, the first cumulant, i.e., the individual spin-$\nu$ ($\nu\in\{+,-\})$ current through junction $\ell$ ($\ell\in{\rm L,R}$), is given by
\begin{align}
\lla  J_\ell^{\nu}\rra =(-i)\frac{\partial}{\partial\chi_{\ell\nu}}\lambda_0(\bm{\chi})|_{\bm{\chi}\to\bm{0}}.
\end{align}
By making use of the quantum master equation (\ref{master_equations_chi_average_spin}), one arrives at
\begin{align}\label{equation_of_polarization}
\lla J_\ell^\pm\rra =-2\kappa_\pm\{&\gamma_{\ell}^+\rho_{00}+{\textstyle \frac{1}{2}}(\tilde{\gamma}_{\ell}^+-\gamma_{\ell}^-)\rho_{11}-\tilde{\gamma}_{\ell}^-\rho_{\rm dd}
\nonumber \\
&\mp(\gamma_{\ell}^-\!\!+\!\tilde{\gamma}_{\ell}^+){\bm S}\!\cdot\!{\bm m}_{\ell}\}.
\end{align}
Apparently, it not only depends on the electron occupation probabilities but also the average spin .
The net charge and spin currents through the junction $\ell$ are defined as
\begin{subequations}\label{Jcs}
\begin{align}
\lla J^{\rm ch}_\ell \rra=\lla J_\ell^+\rra +\lla J_\ell^-\rra,\label{Jcs1}
\\
\lla J^{\rm sp}_\ell \rra=\lla J_\ell^+\rra -\lla J_\ell^-\rra.\label{Jcs2}
\end{align}
\end{subequations}
Specifically, utilizing \Eq{equation_of_polarization}, one finds
\begin{subequations}\label{spin_current_and_current}
\begin{align}
\lla J^{\rm ch}_\ell\rra =&-2[\gamma_{\ell}^+\rho_{00}+{\textstyle \frac{1}{2}}(\tilde{\gamma}_{\ell}^+-\gamma_{\ell}^-)\rho_{11}-\tilde{\gamma}_{\ell}^-\rho_{\rm dd}]
\nonumber \\
&+2p_\ell(\gamma_{\ell}^-+\tilde{\gamma}_{\ell}^+)\boldsymbol{S}\cdot{\bm m}_{\ell},\label{current}
\\
\lla J^{\rm sp}_\ell\rra =&-2p_\ell[\gamma_{\ell}^+\rho_{00}+{\textstyle \frac{1}{2}}(\tilde{\gamma}_{\ell}^+-\gamma_{\ell}^-)\rho_{11}-\tilde{\gamma}_{\ell}^-\rho_{\rm dd}]
\nonumber \\
&+2(\gamma_{\ell}^-+\tilde{\gamma}_{\ell}^+)\boldsymbol{S}\cdot{\bm m}_{\ell}.\label{spin_current}
\end{align}
\end{subequations}

Analogously, the second cumulants, corresponding to the correlations between the spin-$\nu$ currents through junction $\ell$ and
spin-$\nu'$ current through junction $\ell'$, can be obtained via taking the second-order partial derivatives with respect to their corresponding counting fields
\begin{align}\label{C2}
\lla  J_\ell^{\nu}J_{\ell'}^{\nu'}\rra =(-i)^2\frac{\partial^2}{\partial\chi_{\ell\nu}\partial\chi_{\ell'\nu'}}\lambda_0(\bm{\chi})|_{\bm{\chi}\to\bm{0}}.
\end{align}
According to \Eq{Jcs}, the noises of the net charge and spin currents are thus given by
\begin{subequations}\label{Schsp}
\begin{align}
S_{\ell\ell'}^{\rm ch}\!=\!\lla  J_\ell^{+}\!J_{\ell'}^{+}\rra\!+\lla  J_\ell^{-}\!J_{\ell'}^{-}\rra\!+\lla  J_\ell^{+}\!J_{\ell'}^{-}\rra\!+\lla  J_\ell^{-}\!J_{\ell'}^{+}\rra
,
\\
S_{\ell\ell'}^{\rm sp}\!=\!\lla  J_\ell^{+}\!J_{\ell'}^{+}\rra\!+\lla  J_\ell^{-}\!J_{\ell'}^{-}\rra\!-\lla  J_\ell^{+}\!J_{\ell'}^{-}\rra\!-\lla  J_\ell^{-}\!J_{\ell'}^{+}\rra,
\end{align}
\end{subequations}
where each term is evaluated according to \Eq{C2}. Higher order cumulants can be obtained in an analogous manner.
Although there were investigations about charge noise in spin valves, it is still of great importance to investigate the spin-resolved noises, which are essential to reveal the underlying spin correlations in transport.

\section{\label{results}Results and Discussion}

\begin{figure}
\centering
\includegraphics[scale=0.35]{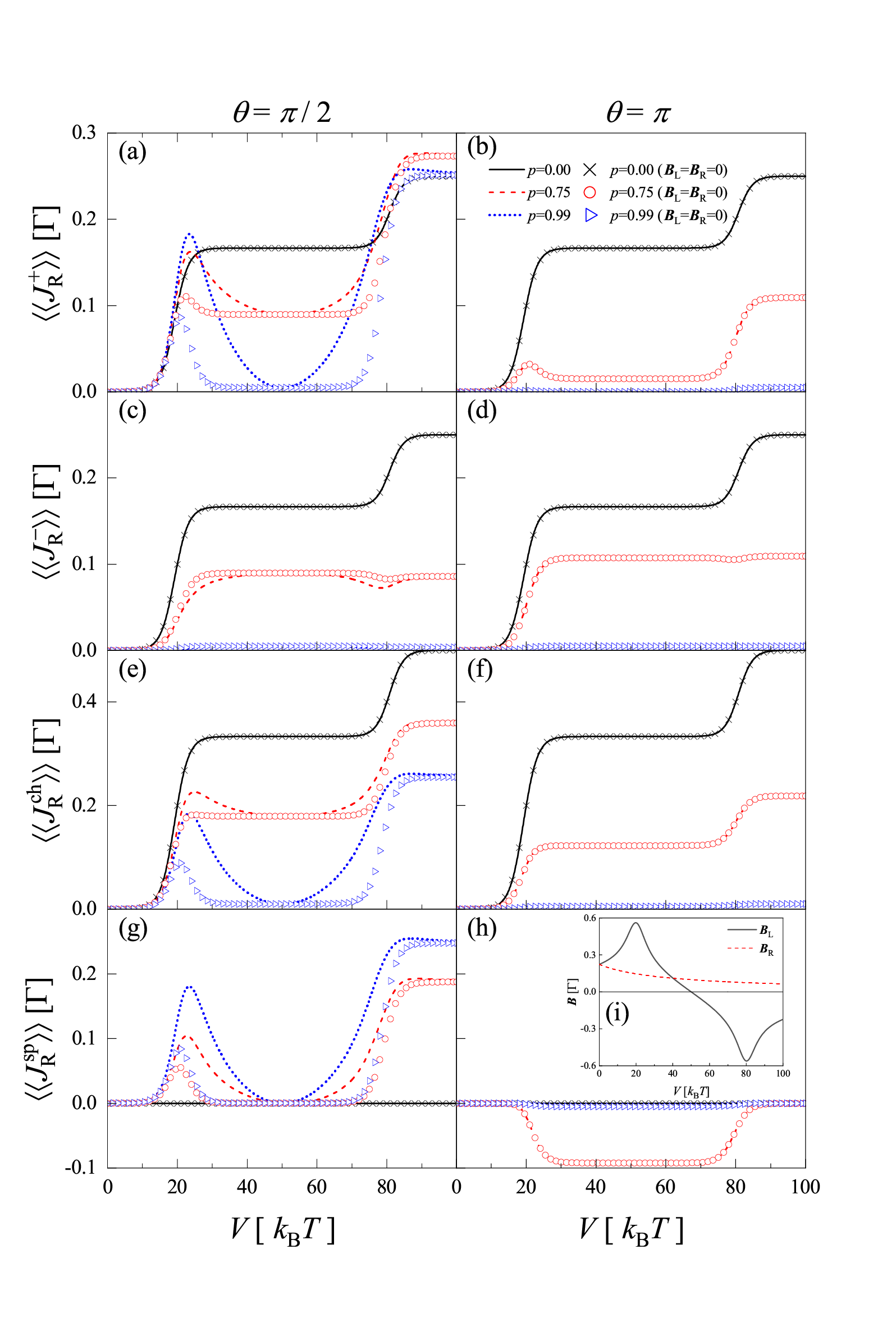}
\caption{\label{fig2} Individual spin-resolved currents ($\lla  J_{\rm R}^{+}\rra$ and $\lla  J_{\rm R}^{-}\rra$), the net charge and spin currents ($\lla  J_{\rm R}^{\rm ch}\rra$ and $\lla  J_{\rm R}^{\rm sp}\rra$) versus bias voltage through the right junction for perpendicularly ($\theta=\frac{\pi}{2}$) and antiparallely ($\theta=\pi$) aligned magnetizations with various polarizations.
For comparison, the results neglecting the exchange magnetic field are also plotted in symbols.
We choose symmetric tunneling couplings $\Gamma_{\rm L}=\Gamma_{\rm R}=\Gamma/2$ and same polarizations $p_{\rm L}=p_{\rm R}=p$. The other parameter are $\varepsilon=10k_{\rm B}T$ and $U=30k_{\rm B}T$.
Insect: The exchange field due to coupling to the left and right electrodes versus bias voltage for full polarization $p_L=p_R=p=1$.}
\end{figure}

\subsection{Current-voltage characteristics}
Utilizing \Eq{master_equations_chi_average_spin} it is easy to check that in the stationary limit $\lla J^{\rm ch}_{\rm L}\rra +\lla J^{\rm ch}_{\rm R}\rra=0$, which ensures the charge conservation.
It is thus enough to analyze either of the currents through the left or right junctions.
In \Fig{fig2}, we plotted the individual spin-resolved currents $\lla J^{\pm}_{\rm R}\rra$, net charge current $\lla J^{\rm ch}_{\rm R}\rra$, and net spin current $\lla J^{\rm sp}_{\rm R}\rra $ through the right junction vs bias voltage for various polarizations with perpendicular aligned ($\theta=\frac{\pi}{2}$) and antiparallel ($\theta=\pi$) electrode magnetizations.

Let us first consider the case of perpendicular alignment ($\theta={\textstyle \frac{\pi}{2}}$).
For nonmagnetic electrodes ($p=0$), the exchange field has a vanishing contribution [cf. \Eq{Bell}], and thus the system can be mapped onto a resonant tunneling model.
At low bias, the transport is blocked due to vanishing occupation of the QD.
As the bias increases, whenever an excitation energy level  ($\varepsilon$ or $\varepsilon+U$) falls into the energy window defined by the chemical potentials of the left and right electrodes, a new transport channel opens. This leads to step-like structures in the individual spin-resolved currents and net charge current, as shown by the solid lines in \Figs{fig2}(a), (c), and (e).
A step occurs at a bias twice of the excitation energy level due to symmetric application of the bias voltage across the QD ($\mu_{\rm L/R}=\pm V/2$).
In this case, $\lla J_{\rm R}^+\rra$ and $\lla J_{\rm R}^-\rra$ are equal in magnitude and the net spin current thus is zero according to \Eq{Jcs2}, cf. the solid line in \Fig{fig2}(g).

Finite spin polarization in the electrodes gives rise to spin accumulation in the QD, which will have essential role to play in transport.
At the first current plateau ($2\varepsilon<V<2(\varepsilon+U)$), double occupation on the QD is energetically not allowed and the tunneling rates are greatly reduced at low temperatures, i.e., $\gamma_{\rm R}^+,\tilde{\gamma}_{\rm R}^+\to0$ and $\gamma_{\rm R}^-,\tilde{\gamma}_{\rm R}^-\to\Gamma_{\rm R}$. As a result, the
spin-resolved currents in \Eq{equation_of_polarization} are simplified to
\begin{align}\label{equation_of_polarization_R}
\lla J_{\rm R}^\pm\rra =\kappa_{\rm R\pm}\Gamma_{\rm R}\rho_{11}\pm \sqrt{2}\kappa_{\rm R\pm}\Gamma_{\rm R}(S_{x}
-S_{y}).
\end{align}
The first term is directly associated with the probability of an electron in the QD to tunnel out via the right junction, where the rates are modulated by the corresponding densities of states of majority or minority spins in the electrode.
The second term is unambiguously related to the spin accumulation on the QD.
In the absence of the exchange field (${\bm B}=0$), the occupation and average spin on the QD in the steady state can be obtained by utilizing \Eqs{master_equations_chi_average_spin}:
\bsube\label{elepi1st}
\be
\rho_{11}= \frac{2\Gamma_{\rm L}}{2\Gamma_{\rm L}+(1-p^2)\Gamma_{\rm R}},
\ee
and
\bea\label{Sxy1stpl}
S_x&=&-\frac{p^3\Gamma_{\rm R}}{\sqrt{2}(2\Gamma_{\rm L}+(1-p^2)\Gamma_{\rm R})},
\\
S_y&=&\frac{p(2-p^2)\Gamma_{\rm R}}{\sqrt{2}(2\Gamma_{\rm L}+(1-p^2)\Gamma_{\rm R})}.
\eea
\esube
As polarization $p$ increases, the QD is inclined to be occupied by one extra electron ($\rho_{11}\to1$) and the average spins  $S_x\to-\sqrt{2}\Gamma_{\rm R}/(4\Gamma_{\rm L})$ and $S_y\to\sqrt{2}\Gamma_{\rm R}/(4\Gamma_{\rm L})$ such that the total spin tends to align along $-{\bm m}_{\rm R}$, i.e., antiparallel to the magnetization of the right electrode (see \Fig{fig1}).
Due to the presence of Coulomb repulsion, no more electrons can tunnel into the QD and thus transport is strongly suppressed.
This explains the strong reduction of the spin-resolved currents with rising $p$, see the dotted curves in \Figs{fig2}(a) and (c).

The presence of exchange field leads to a precession of the average spin. It is shown in the inset of \Fig{fig2}(h) that ${\bm B}_{\rm L}$ plays the dominant role, especially when the bias is in resonance with the energy levels $\varepsilon$ and $\varepsilon+U$.
The average spin now rotates dominantly about ${\bm B}_{\rm L}$ and thus weakens the spin valve effect.
This increases the probability for an electron to leave the dot and leads to an enhanced spin-resolved current $J_{\rm R}^{+}$, particularly prominent at $V=2\varepsilon$ and $V=2(\varepsilon+U)$.
This also explains the strong enhancement of the net spin current $\lla J_{\rm R}^{\rm sp}\rra$ at $V=2\varepsilon$, see for instance the dotted curve in \Fig{fig2}(g).
However, the suppression of $\lla J_{\rm R}^{-}\rra$ is not lifted due to strongly annihilated $\kappa_{\rm R-}$ in the limit of large polarization.

As the bias further increases to the regime $V>2(\varepsilon+U)$, double occupation on the QD is now energetically allowed. This opens up an additional transport channel and the spin-resolved currents rise to the second plateau, cf. the solid curves in \Figs{fig2}(a) and (c) for $p=0$.
For bias far away from the energy level $\varepsilon+U$, the influence of the exchange field is greatly reduced even for a strong polarization (see the inset of \Fig{fig2}(h)).
At low temperatures, the tunneling rates in \Eq{equation_of_polarization} are well approximated by either 0 or $\Gamma_{\rm R}$.
One obtains simple expressions of the spin-resolved currents for finite spin polarization:
\be\label{JRPM}
\lla J_{\rm R}^\pm\rra =\kappa_{\rm R\pm} \Gamma_{\rm R}(\rho_{11}+2\rho_{\rm dd})\pm  \sqrt{2}  \kappa_{\rm R\pm} \Gamma_{\rm R}(S_{x} -S_{y}).
\ee
As the polarization increases, $\kappa_{\rm R-}$ is strongly reduced, leading thus to a prominent suppression of $\lla J_{\rm R}^{-}\rra $, in comparison with that of $p=0$, cf. \Fig{fig2}(c).
Remarkably, finite polarization gives rise to a slight increase of $\lla J_{\rm R}^{+} \rra$.
For symmetric tunnel couplings ($\Gamma_{\rm L}=\Gamma_{\rm R}$) that we considered here, utilizing \Eq{master_equations_chi_average_spin} one finds the stationary probabilities and average spin
\begin{gather}
	\rho_{11}+2\rho_{\rm dd}\to 1,\;  S_x\to 0,\; S_y\to-\sqrt{2}p/4.
\end{gather}
Thus, there is a competition between the first and second term in \Eq{JRPM}.
For $p=0$, the first term dominates, which gives the second current plateau, cf. the solid curve in \Fig{fig2}(a).
An increase of $p$ leads to a rising $\kappa_{\rm R+}$ but a decreasing $S_y$, which explains the slight enhancement of $\lla J_{\rm R}^{+}\rra$ in comparison with that of $p=0$, as shown by the dashed and dotted curves in \Fig{fig2}(a).


\begin{figure}
\centering
\includegraphics[scale=0.45]{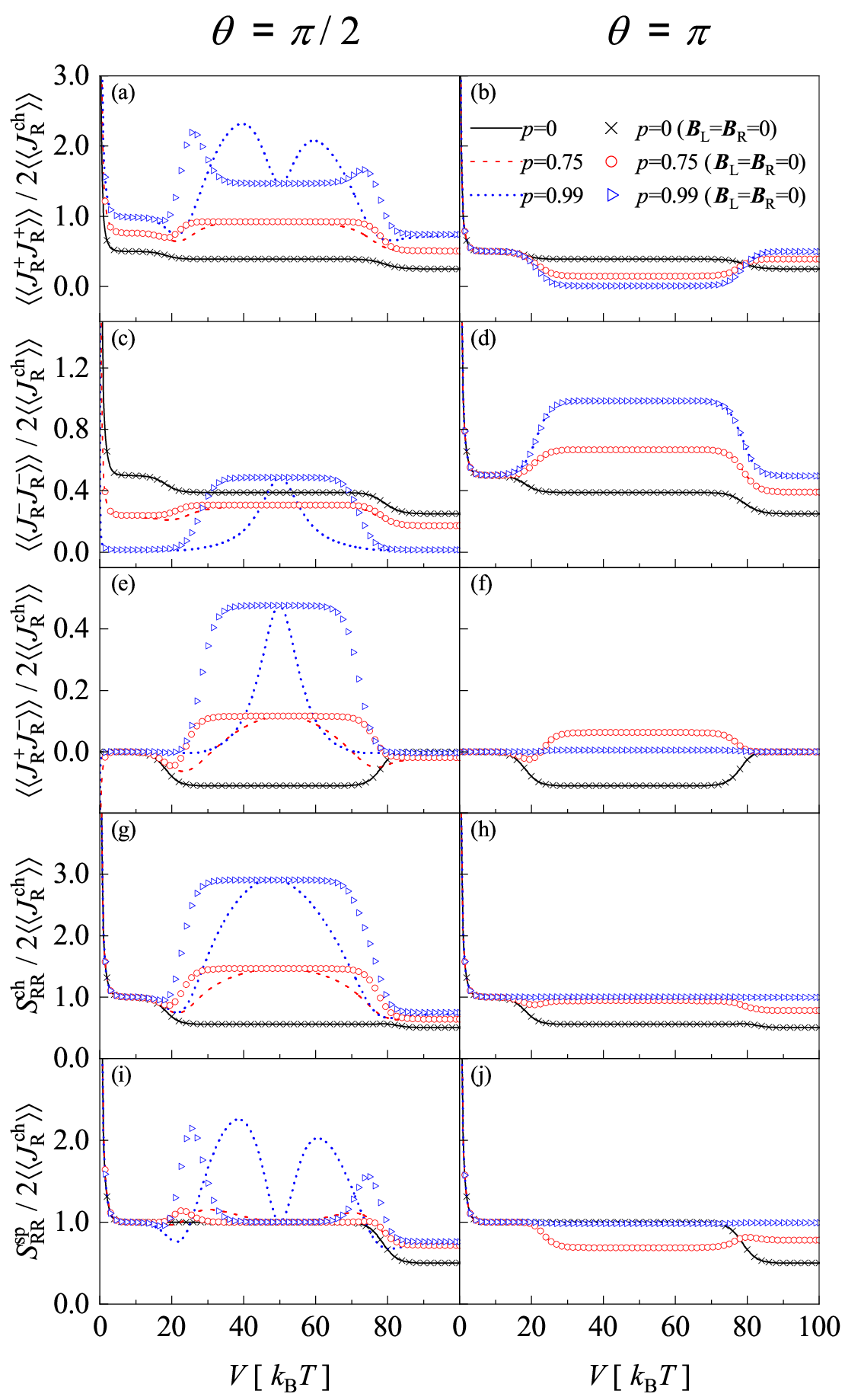}
\caption{\label{fig4}Spin autocorrelations ($\lla J_{\rm R}^+J_{\rm R}^+\rra$ and $\lla J_{\rm R}^-J_{\rm R}^-\rra$), cross-correlation ($\lla J_{\rm R}^+J_{\rm R}^-\rra$), and noises of net charge and spin currents ($S_{\rm RR}^{\rm ch}$ and $S_{\rm RR}^{\rm sp}$) versus bias voltage for perpendicular ($\theta=\frac{\pi}{2}$) and antiparallel ($\theta=\pi$) magnetization alignments with various polarizations.
The results neglecting the exchange magnetic field are also plotted in symbols for comparison.
The other parameters are the same as those in \Fig{fig2}.}
\end{figure}

Let us now consider the situation for the antiparallel configuration ($\theta=\pi$).
For $p=0$, the charge and spin currents increase with bias in a typical step-like manner, the same as those for perpendicular configuration ($\theta=\frac{\pi}{2}$).
In the case of large polarization and in the bias region $2\varepsilon<V<2(\varepsilon+U)$, whenever a spin ``$+$'' electron tunnels into the QD, it will be trapped in the QD for a long time. Its average spin is found to be along ${\bm m}_{\rm L}$, almost the same direction of the exchange field (see \Eq{Bell} and the inset of \Fig{fig2}(h)), which can not weaken the spin valve effect.
Furthermore, the Coulomb interaction energetically prohibits a second electron to tunnel into QD and transport is thus strikingly suppressed with increasing polarization.
For a large bias $V>2(\varepsilon+U)$, although double occupation is allowed, the probability for a second electron of spin ``$-$'' tunneling through the QD is strongly reduced  due to vanishing density of state for ``$-$'' spin in the left electrode.
One thus observes prominent suppression of spin-resolved currents as well as net charge and spin currents, as shown by the dotted curves in \Fig{fig2}(b), (d), (f), and (h).

\subsection{Spin current noise characteristics}

Now we are in a position to investigate the spin-resolved current noises based on \Eq{C2}.
Again, we consider two different magnetization configurations, i.e., perpendicular ($\theta=\frac{\pi}{2}$) and antiparallel ($\theta=\pi$) alignments.
The numerical results $\lla J_{\rm R}^{\nu}J_{\rm R}^{\nu'}\rra$ are presented in \Fig{fig4}, where noises measured in terms of the Fano factors are plotted as functions of the bias voltage.
The noises between different electrodes are quantitatively similar.
Furthermore, the spin cross-correlations satisfy $\lla J_{\rm R}^{+}J_{\rm R}^{-}\rra=\lla J_{\rm R}^{-}J_{\rm R}^{+}\rra$ such that one only needs to consider one of them.

Let us first consider the situation of perpendicular alignment ($\theta=\frac{\pi}{2}$).
At low bias ($V\ll k_{\rm B}T$), the thermal noise dominates, which is described by the well-known hyperbolic cotangent behavior. This leads to divergent spin auto-correlations ($\lla J_{\rm R}^{\nu}J_{\rm R}^{\nu}\rra/2\lla J^{\rm ch}_{\rm R}\rra \to \infty, \nu=+,-$) for various polarizations, as shown in \Figs{fig4}(a) and (c).
The spin ``$+$'' and spin ``$-$'' currents are found to be uncorrelated for $p=0$, see the solid curve in \Fig{fig4}(e).
Yet, finite polarization gives rise to a negative cross-correlation ($\lla J_{\rm R}^{+}J_{\rm R}^{-}\rra/2\lla J^{\rm ch}_{\rm R}\rra$), as shown by the dashed and dotted curves in \Fig{fig4}(e).

As bias increases but remains lower than the first excitation energy level ($V<2\varepsilon$), electron transport is exponentially suppressed. Charge tunneling events are uncorrelated and thus both noises of net charge current and spin current exhibit Poissonian statistics independent of polarizations $(S^{\rm ch}_{\rm RR}/2\lla J^{\rm ch}_{\rm R}\rra = 1$ and $S^{\rm sp}_{\rm RR}/2\lla J^{\rm ch}_{\rm R}\rra = 1)$, as shown in \Figs{fig4}(g) and (i), respectively.
Remarkably, the spin autocorrelations depend sensitively on the polarizations: $S^{++}_{\rm RR}$ increases but $S^{--}_{\rm RR}$ decreases with rising $p$.
This demonstrates that the autocorrelations of spin-resolved currents may serve as a skeptical tool to detect degree of spin polarization in FM materials.

As the bias further increases, the first and then the second channels open. Both $\lla J_{\rm R}^{+}J_{\rm R}^{+}\rra$ and $\lla J_{\rm R}^{-}J_{\rm R}^{-}\rra$ decreases in a step-like manner for $p=0$.
As polarization increases, a spin ``$+$'' electron tunneled into the QD tends to stay there for a long time due to the spin valve effect.
When it is tunneled out, a bunching of ``$+$'' spin electrons can flow during a short time window, leading thus to a dynamical spin blockade mechanism \cite{Mut21245412,Kon21155414,Per07193301}. This explains the strongly enhanced super-Poissonian spin autocorrelation ($\lla J_{\rm R}^{+}J_{\rm R}^{+}\rra/2\lla J^{\rm ch}_{\rm R}\rra > 1 $) in the bias regime $2\varepsilon <V< 2(\varepsilon+U)$, cf. the dotted curve in \Fig{fig4}(a).
The existence of the exchange field weakens the spin-valve effect and consequently reduces the noise, leading thus to a double-peak structure in $\lla J_{\rm R}^{+}J_{\rm R}^{+}\rra$.
The cross-correlation between the ``$+$'' and ``$-$'' spin currents ($\lla J_{\rm R}^{+}J_{\rm R}^{-}\rra$) is also sensitive to the polarization: It  changes its sign from negative to positive as $p$ increases.
The noise characteristics of the net charge current ($S^{\rm ch}_{\rm RR}$) and net spin current ($S^{\rm sp}_{\rm RR}$) thus can be understood in terms of the individual components according to \Eq{Schsp}.
We remark that the unique double-peak structure in the noise of the net spin current may serve as a sensitive means to measure the exchange magnetic field.

\begin{figure}
\centering
\includegraphics[scale=0.3]{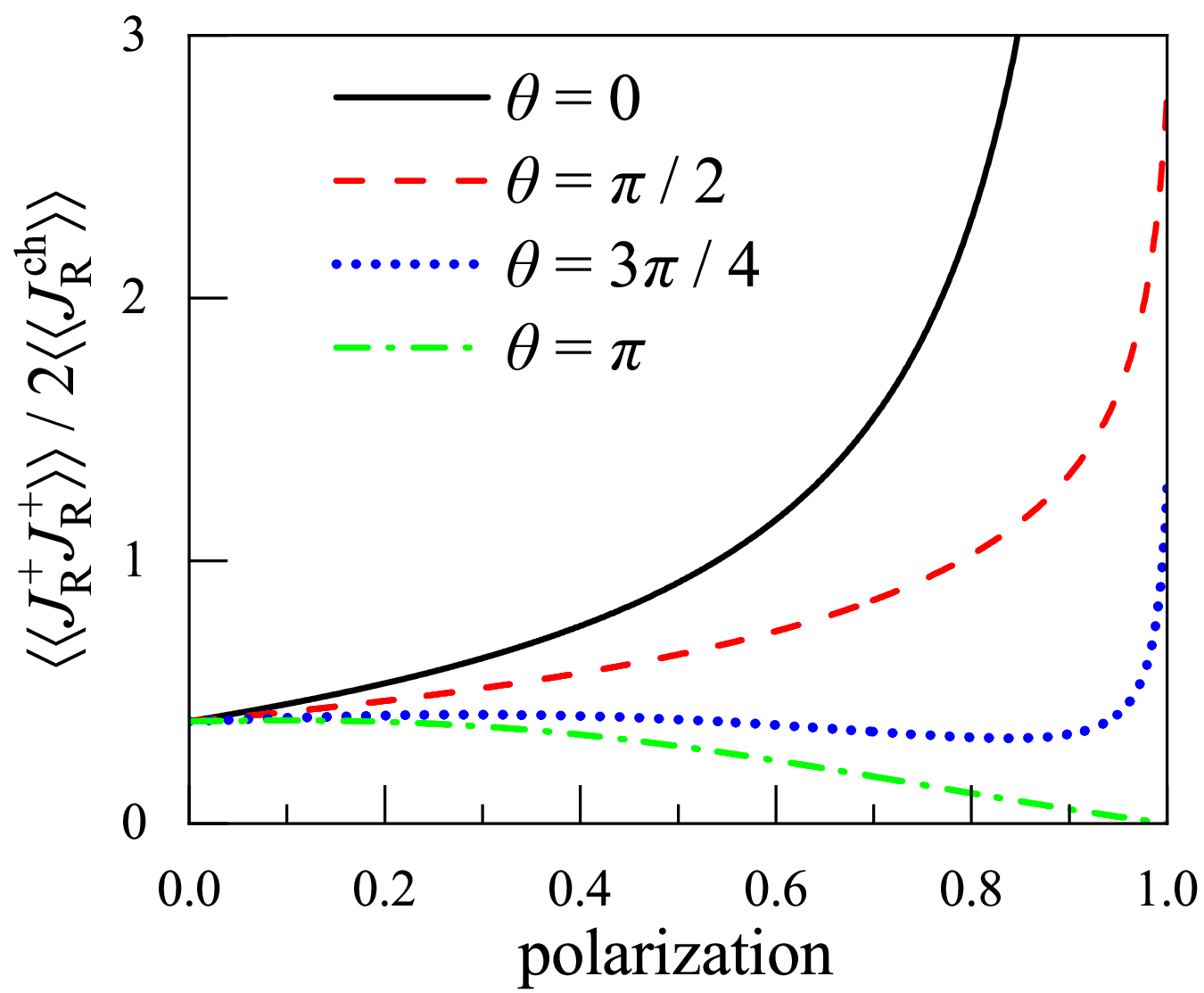}
\caption{\label{fig5}Spin current autocorrelation $\lla J_{\rm R}^{+}J_{\rm R}^{+}\rra$ versus polarization for various magnetization alignments at a particular bias voltage $V=40{k_{\rm B}T}$ .
The other parameter are the same as those in \Fig{fig2}.}
\end{figure}

For the antiparallel configuration ($\theta=\pi$) and in the low bias regime ($V\ll k_{\rm B}T$), current is suppressed and the noises are quantitatively similar to those for the perpendicular configuration.
As bias increases, the noises show typical step-like structure.
In comparison with the situation of perpendicular alignment, the noises for $\theta=\pi$  show distinct behaviors.
First, the noises for antiparallel configuration is insensitive to the exchange filed. This is due to the fact that in this case the accumulated spin is almost in the same direction of ${\bm m}_{\rm L}$ such that the exchange field has a vanishing role to play.
Second, for bias $V<2\varepsilon$, both spin autocorrelations and cross-correlations are independent of the polarizations ($\lla J_{\rm R}^{\mu}J_{\rm R}^{\nu}\rra\rra\to\frac{1}{2}$ and $\lla J_{\rm R}^{+}J_{\rm R}^{-}\rra\to0$), implying that in this case the spin ``$+$'' and ``$-$'' currents are uncorrelated for an arbitrary polarization.
Third, in the bias regime $2\varepsilon <V< 2(\varepsilon+U)$, the autocorrelation $\lla J_{\rm R}^{+}J_{\rm R}^{+}\rra$ shows opposite dependence on polarization in comparison to the perpendicular configuration.
For a deep analysis, we plotted in \Fig{fig5} the autocorrelation $\lla J_{\rm R}^{+}J_{\rm R}^{+}\rra$ versus $p$ for various $\theta$ with a given bias voltage $V=40{k_{\rm B}T}$ inside the first current plateau.
As $p\to0$, $\lla J_{\rm R}^{+}J_{\rm R}^{+}\rra$ for different alignments are consistent.
For $\theta=0$ or $\frac{\pi}{2}$, $\lla J_{\rm R}^{+}J_{\rm R}^{+}\rra$ rises monotonically with polarization, leading to prominent super-Possionian noises at a large polarization, see the solid and dashed curves.
For $\theta=\frac{3}{4}\pi$, $\lla J_{\rm R}^{+}J_{\rm R}^{+}\rra$ first decreases for a wide range of $p$ and then increases rapidly in the limit of $p\to1$.
For antiparallel configuration ($\theta=\pi$), $\lla J_{\rm R}^{+}J_{\rm R}^{+}\rra$ decreases  monotonically with polarization and vanishes at $p=0$.
We remark that this unique noise feature may serve as a sensitive tool to measure the magnetization directions between the left and right FM electrodes.

\section{\label{conclusion}Conclusion}

We have investigated the spin-dependent transport through a spin valve composed of a quantum dot tunnel coupled to external ferromagnetic electrodes with noncollinear magnetizations.
The analysis is based on the spin-resolved full counting statistics, which allowed us to determine systematically spin-resolved transport characteristics.
In particular, we have analyzed individual spin-resolved currents, as well as their auto- and cross-correlations versus bias for different magnetization configurations and polarizations.
In the case of perpendicular alignment, we found that the charges are transferred uncorrelated at a low bias; however, the autocorrelations of spin currents were shown to be sensitive to the polarization in the electrodes.
As bias increases to the regime where single occupation of electron is allowed, the interplay between tunnel coupling and the Coulomb interaction gives rise to a prominent exchange field. It leads to the precession of the accumulated spin in QD, which lifts the bunching of spin tunneling events and thus results in a unique double-peak structure in the noise of the net spin current. This distinct feature may potentially be used to measure the exchange field.
Furthermore, we have revealed that the spin autocorrelation undergoes a radical change for different FM alignments in the limit of large polarization, which may be utilized as a sensitive tool to measure the magnetization directions between the ferromagnetic electrodes.
Our results demonstrated unambiguously the superiority of spin-resolved noise correlations in revealing underlying spin transport dynamics and characteristics.

\begin{acknowledgments}
This work is supported by the National Natural Science Foundation of China (Grant Nos. 11774311 and 12005188) and education department of Zhejiang Province.
\end{acknowledgments}


\end{document}